\documentclass[aps,
 jmp,%
 amsmath,amssymb,
reprint,superscriptaddress,twoside,twocolumn,floatfix,a4paper,longbibliography]{revtex4-1}

\usepackage{graphicx}
\usepackage{amsmath}
\usepackage{dcolumn}   % needed for some tables
\usepackage{bm}        % for math
\usepackage{amssymb}   % for math
\usepackage{graphicx,epsfig}
\usepackage{color}
\usepackage[usenames,dvipsnames]{xcolor}
\usepackage[ocgcolorlinks,colorlinks=true,linkcolor=blue,citecolor=red]{hyperref}
\usepackage{amsmath,amssymb, amsthm, amsfonts}
\usepackage{xspace}
\usepackage{xcolor}
\usepackage{verbatim}
\usepackage{mathtools}

\usepackage{natbib}
\usepackage{mathrsfs}
\usepackage{amsmath}
\usepackage{mathtools}
\usepackage{graphicx}
\usepackage{float}

\begin{document}

\title{Spontaneous symmetry breaking-induced thermospin effect in superconducting tunnel junctions}
\author{Gaia Germanese}
\affiliation{Dipartimento di Fisica dell’Università di Pisa, Largo Pontecorvo 3, I-56127 Pisa, Italy}
\affiliation{NEST Istituto Nanoscienze-CNR and Scuola Normale Superiore, I-56127 Pisa, Italy}
\author{Federico Paolucci}
\affiliation{NEST Istituto Nanoscienze-CNR and Scuola Normale Superiore, I-56127 Pisa, Italy}
\author{Giampiero Marchegiani}
\affiliation{Quantum Research Centre, Technology Innovation Institute, Abu Dhabi, UAE}
\author{Alessandro Braggio}
\affiliation{NEST Istituto Nanoscienze-CNR and Scuola Normale Superiore, I-56127 Pisa, Italy}
\author{Francesco Giazotto}
\affiliation{NEST Istituto Nanoscienze-CNR and Scuola Normale Superiore, I-56127 Pisa, Italy}

\begin{abstract}
We discuss the charge and the spin tunneling currents between two Bardeen-Cooper-Schrieffer (BCS) superconductors, where one density of states is spin-split by the proximity of a ferromagnetic insulator. In the presence of a large temperature bias across the junction, we predict the generation of a spin-polarized thermoelectric current. This thermo-spin effect is the result of a spontaneous particle-hole symmetry breaking in the absence of any polarizing tunnel barrier. The two spin components, which move in opposite directions, generate a spin current larger than the purely polarized case when the thermo-active component dominates over the dissipative one.
\end{abstract}

\maketitle
\begin{figure}[htp]
\includegraphics[width=1\columnwidth]{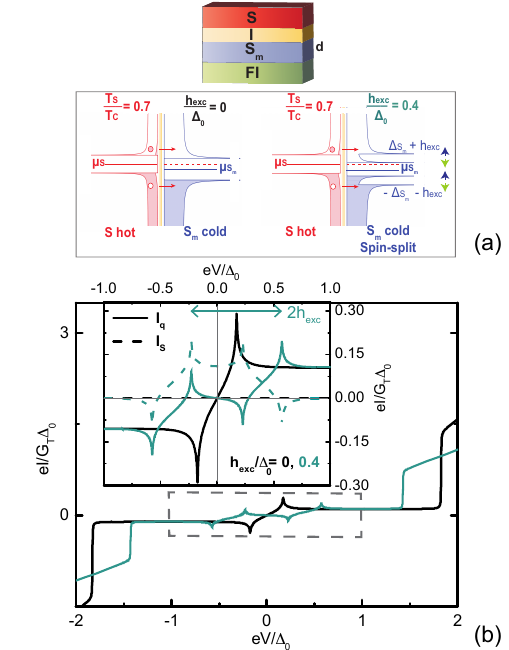}
\caption{(a) Top: $S-I-S_m$ junction building blocks. The top hot [bottom cold] superconductor is indicated with $S$ (${S_m}$), and the exchange field is present in $S_m$ only. Bottom: energy band diagrams of the superconductors are shown in the absence ($h_{\rm{exc}}=0$, left panel) and presence ($h_{\rm{exc}}=0.4\Delta_0$, right panel) of an exchange field ($h_{\rm{exc}}$). For $h_{\rm{exc}}=0$, the DoS of {$S_m$} exhibits two peaks. For $h_{\rm{exc}}=0.4\Delta_0$, the spin-split DoS (blue $\uparrow$ and green $\downarrow$ for the up and down spin components, respectively) shows four peaks. The difference between the chemical potential of the two leads is $\mu_{S}-\mu_{S_m}=eV$. The temperature of the $S$ (${S_m}$) superconductor is set at $T_{S}= 0.7T_{c}$ ($T_{{S_m}}= 0.01T_{c}$). (b) Quasiparticle current-voltage characteristics for $h_{\rm{exc}}=0$ (black) and $h_{\rm{exc}}=0.4\Delta_0$ (aquamarine) display two and four matching peaks, respectively. Inset: Blow-up of the charge ($I_q$, solid lines) and spin currents ($I_s$, dashed lines) at low bias voltage.}
\label{Fig1}
\end{figure}
%%%%%%%%%%%%%%%%%%%%%%%%%%%%%%%%%%%%
%%%%%%%%%%%%%%%%%%%%%%%%%%%%%%%%%%%%%
%%%%%%%%%%%%%%%%%%%%%%%%%%%%%%%%%%%%%%%%%%
%%%%%%%%%%%%%%%%%%%%%%%%%%%%%%%%%%%%%%%%%%
\section{Introduction}
Spintronics consists of the active manipulation of the spin degree of freedom to develop a wide range of solid-state technologies~\citep{ReviewSpintronics, ReviewLinder}. In this context, hybrid systems involving ferromagnets and superconductors have been exploited for the generation of spin-polarized currents~\citep{TedrowPRL27, MeserveyPRB11, TakahashiPRL, Jedema, TakahashiJAP, Yamashita, Maekawa,  GiazottoTaddeiPRB} with applications as memory elements~\citep{HaoPRB, GiazottoF, BergeretVerso, DeSimoni}, due to a nearly perfect spin-valve effect. At the same time, thermoelectricity in these hybrid devices has been discussed theoretically~\citep{Virtanen, Titov, Jacquod, Kalenkov, Heidrich} and experimentally \citep{Eom, Parsons, Jiang, GiazottoPRA, Tan} with intriguing results at non-local level~\citep{Pauli, Dolgirev, Hussein, Kirsanov, BlasiPRL, BlasiPRB, KalenkovPRB2020} and applications as detectors \cite{Heikkila}. 
Yet, in the presence of temperature gradients, the interplay of the magnetic field with the superconducting order parameter gives rise to exotic non-equilibrium phenomena, such as the generation of pure spin currents (spin-Seebeck effect) \citep{Machon, Ozaeta, Linder, BergeretColloquium,  KolendaPRB, Shapiro}. Particle-hole (PH) symmetry breaking is a necessary requirement to generate thermoelectricity in the linear regime, i.e., $\delta V, \delta T\to 0$~\cite{Sanchez, Benenti}. In particular, the violation of this symmetry in ferromagnet-superconductor systems was theoretically demonstrated in the presence of magnetic impurities, which strongly enhance the thermoelectric coefficient \citep{KalenkovPRL, KalenkovPRB2014}.\\
This limitation can be overcome in the presence of a large temperature bias (nonlinear regime), as recently demonstrated in tunnel junctions between superconductors with different energy gaps~\citep{MarchegianiPRL, MarchegianiPRB}.\\
Here, we investigate the thermo-spin effect induced by spontaneous PH symmetry breaking in hybrid ferromagnetic insulator/superconductor systems for a large temperature bias. The effect shares some similarities with the physics reported in Ref.~\cite{MarchegianiPRL}, such as the bipolar thermoelectric nature. On the other hand, it is very different in character since the spin-splitting generates a spin current and charge thermoelectricity even for superconductors of identical gaps. In these conditions, counter-intuitively, dissipative and thermo-active opposite spin components coexist, resulting in maximal spin current efficiency and thermoelectric power generation. We consider a thermally-biased heterostructure of tunnel junctions [see Fig.~\ref{Fig1}(a)]. The system is composed by a Bardeen-Cooper-Schrieffer (BCS) superconductor ($S$) coupled with a ferromagnetic superconductor ($S_m$) by an insulating barrier ($I$). The magnetization of $S_m$ is induced by an exchange interaction due to the proximity of a ferromagnetic insulator ($FI$) when the thickness of $S_m$ ($d$) is smaller than the coherence lenght~\citep{Tokuyasu, Hao}.
This geometry has been realized in recent experiments~\citep{Strambini,StrambiniPRREuS}.  In this system, the exchange interaction ($h_{\rm{exc}}$) breaks the degeneracy between the spin up ($\uparrow$) and spin down ($\downarrow$) components of the superconducting Density of States (DoS)~\footnote{We assume the Josephson coupling can be suppressed via Franhoufer interference, or SQUID interferometry, by applying an in-plane or out-of-plane magnetic field, respectively~\cite {MarchegianiPhase, Barone}.}.\\
We first give an intuitive description of the effect by analyzing the charge current ($I_q$) and the spin current ($I_s$) in the presence of a thermal bias, i.e., $T_S > T_{S_m}$, where $T_S~(T_{S_m})$ is the quasiparticle temperature of $S$ ($S_m$). Figure~\ref{Fig1}(b) shows $I_q$ (solid curves) as a function of the bias voltage  ($V$) in the absence [black, see also left energy diagram in the box of Fig.~\ref{Fig1}(a)], and in the presence [aquamarine, see also right energy diagram in the box of Fig.~\ref{Fig1}(a)] of a sizeable $h_{\rm exc}$. In both cases, $I_q$ is linear in the voltage bias ($I_q\propto V$) for large $V$ (Ohmic response). For low values of $V$, the trace is highly nonlinear and current peaks appear, corresponding to the matching between the singularities in the superconducting DoS~\cite{ShapiroIBM}. Note that each of the two antisymmetric peaks in the $I_q(V)$ characteristics for $h_{\rm exc}=0$ is doubled for $h_{\rm exc}\neq 0$, since the spin-degeneracy is broken by the exchange interaction. In the inset, we magnify the low-voltage behaviour (dashed-rectangle), and we include the corresponding spin current (dashed lines). Clearly, $I_s$ is exactly zero for $h_{\rm exc}= 0$, and finite for $h_{\rm exc}\neq 0$. Moreover, $I_q$ flows against the bias voltage for small values of $V$ when $h_{\rm exc} \neq 0$, i.e., the junction produces thermoelectric power, while it is always dissipative for $h_{\rm exc}=0$. When the active thermo-spin current is generated, it results $|I_s|>|I_q|$. This inequality is never realized in a perfect spin-polarized barrier (i.e. $100\%$ polarization), where $|I_s|=|I_q|$~\cite{ReviewSpintronics}. Indeed, in our structure, the two finite spin-current components (${I_{\uparrow}}$ and ${I_{\downarrow}}$) flow in opposite directions. Moreover, when the thermo-active component dominates, a net thermoelectric spin-polarized current occurs.\\
%%%%%%%%%%%%%%%%%%%%%%%%%%%%%%%%%%%%%%%%%%%%%%%%%%%%%%%%%%%%%%%%%
\section{Model}
We assume $S_m$ to be much thinner than its superconducting coherence length ($d \ll \xi_{0,m}$), in order to consider a homogeneous exchange interaction in $S_m$~\cite{Tokuyasu}. The normalized DoS of the spin-$\sigma$ component [with $\sigma=+(\uparrow),-(\downarrow)$] for the $\alpha$ superconducting electrode (with $\alpha=S, S_m$) is expressed as
\begin{equation}
N_{\alpha \sigma}(E_\alpha)= \left |\Re \!\! \left[\frac{E_{\alpha}+ \sigma h_{\alpha} + i\Gamma}{\sqrt{(E_{\alpha}+\sigma h_{\alpha}+ i\Gamma)^2-{\Delta_{\alpha}}^2}}\right]\!\!\right |,
\label{Eq:1DOS}
\end{equation}
where $h_\alpha$ is the exchange field, $E_{\alpha}=E-\mu_{\alpha}$ is the quasiparticle energy measured with respect to the chemical potential $\mu_\alpha$~\footnote{For simplicity, we assume that the chemical potential for the two spin-species is equal $\mu_\uparrow=\mu_\downarrow$ (no spin accumulation). This approximation holds when the spin-relaxation rate in each electrode is larger than the spin-injection one. The latter scales with spin current of the junction as $I_s\propto G_T$, i.e. for sufficiently opaque tunneling barrier.}, $\Delta_{\alpha}$ is the self-consistent superconducting energy gap for the $\alpha$-lead~\footnote{The self-consistency relation is~\citep{GiazottoPRA, Sarma}:
\begin{equation}
\log\left(\frac{\Delta_0}{\Delta_\alpha}\right)=\int_{0}^{\hbar \omega_D}dE \,\frac{f_+(E)-f_-(E)}{\sqrt{E^2+\Delta_\alpha^2}},
\end{equation}
where $\Delta_0$ is the zero-temperature pairing potential in absence of the exchange field, $\omega_D$ is the Debye frequency and $f_{\pm}(E)=(1+\exp[(\sqrt{E^2+ \Delta ^2}\mp h_{\rm exc})/k_B T_{R}])^{-1}$ with $k_B$ Boltzmann constant.}, and $\Gamma$ is the phenomenological Dynes parameter \cite{Dynes} \footnote{The value of the parameter used for the calculation is $ \Gamma = 10^{-3} \Delta_0$, consistent with the state-of-the-art nanofabrication results.~\cite{StrambiniPRREuS} A more detailed study of the dependence with $\Gamma$ can be also found in \cite{Supp}.}. The sum of the two spin contributions gives the total DoS of a spin-split superconductor $N_{\alpha}(E)=\sum_{\sigma=\pm}N_{\alpha\sigma}(E)/2$. We will adopt the usual approximations found to be valid in many experiments~\footnote{ For simplicity, we neglect the spin-flip and spin-orbit effects, which are not universal and material and geometry dependent \citep{KolendaPRB, Strambini, DeSimoni, StrambiniPRREuS}.}.
For $\alpha=S$, we assume $h_S=0$. Note that, even in the presence of an exchange field, the DoS is PH symmetric, i.e., $N_\alpha(E_\alpha)=N_\alpha(-E_\alpha)$ \cite{Tinkham2004}, and spin component satisfies $N_{\alpha\sigma}(E_\alpha)=N_{\alpha\bar{\sigma}}(-E_\alpha)$, with $\bar{\sigma}=-\sigma$.\\
The quasiparticle current of the spin-$\sigma$ component reads
\begin{equation}
I_{\sigma}= \frac{G_T}{2e}\!\!\int_{-\infty}^{\infty}\!\!\!\!\!\!\!\!dE~N_{S\sigma}(E-eV) N_{S_m\sigma}(E) F_{SS_m}(E),
\end{equation}
where $e$ is the electron charge, $G_T$ is the normal-state conductance of the tunnel junction \footnote{ $G_T=G_\uparrow + G_\downarrow$ and we assume the two spin to have equally contribute $G_\uparrow=G_\downarrow$.}, and $F_{S S_m}(E)=f_S(E-eV)-f_{S_m}(E)$ is the difference between the Fermi-Dirac quasiparticle distributions of two electrodes $f_\alpha(E,T_\alpha)=[1+\exp(E/k_B T_\alpha)]^{-1}$ with $\alpha=S,S_m$. We assume to work in quasi-equilibrium regime, where each electrode is separately at the thermal equilibrium and the electronic temperature can differ from the phononic one, as experimentally demonstrated \citep{GiazottoReview, MuhonenReview, GiazottoNature12, FornieriReview}. 
The charge current and the spin current are defined as $I_q = I_{\uparrow}+I_{\downarrow}$ and $I_s = I_{\uparrow}-I_{\downarrow}$, respectively. By exploiting the PH symmetry, we note that the charge current is odd (even) in $V$ ($h_{\rm exc}$) with $I_q(V, h_{\rm exc})=-I_q(-V, h_{\rm exc})=I_q(V, -h_{\rm exc})$, while  the spin current is an even (odd) function in $V$ ($h_{\rm exc}$) with $I_s(V, h_{\rm exc})=I_s(-V, h_{\rm exc})=-I_s(V, -h_{\rm exc})$, as shown in Fig.~\ref{Fig1}(b)\cite{Supp}.\\
We assume the two superconductors ($S$ and $S_m$) to have the same zero-temperature energy gap ($\Delta_{S,0}=\Delta_{S_m,0}=\Delta_0$ and, hence, the same critical temperature $T_c$). Therefore, no thermo-electric effect occurs for $h_{\rm exc}=0$~\citep{MarchegianiPRL, MarchegianiPRB}.\\ 
%%%%%%%%%%%%%%%%%%%%%%%%%%%%%%%%%%%%%%%%%
%%%%%%%%%%%%%%%%%%%%%%%%%%%%%%%%%%%%%%
\section{Charge thermoelectric effect}
\begin{figure}[tbp]
\includegraphics[width=1\columnwidth]{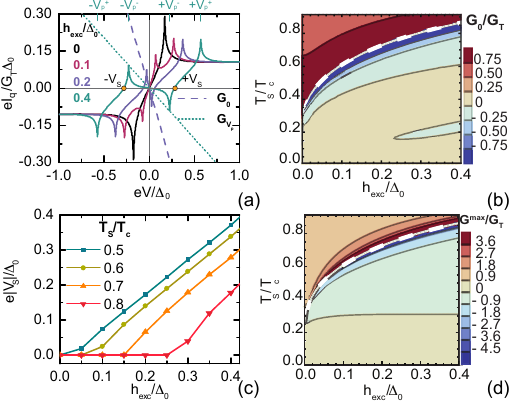}
\caption{(a) Quasiparticle current-voltage $I_q(V)$ characteristics are shown for different values of $h_{\rm exc}/\Delta_0$. In the linear regime, the IV curves can be approximated by the linear $I_q=G_0 V$ of Eq.~\eqref{Eq:G0} (dashed violet line), while, in the nonlinear regime in bias, the $G^{max}$ evaluated at the thermoelectric peaks (dotted aquamarine line). The Seebeck voltages ($\pm V_S$) (orange circles) and the matching peaks values ($\pm V_p^{\pm}$) are reported for $ h_{ex}=0.4\Delta_0$. (b) In the linear regime, the zero-bias conductance ($G_0$) as a function of $h_{\rm exc}$ and $T_S$ is shown, distinguishing thermo-active areas (blue tones) from dissipative ones (red tones). (c) Absolute value of the Seebeck voltage as a function of $h_{\rm exc}$ is shown for different values of the thermal bias ($T_S$). (d) Maximum conductance evaluated at the matching peak ${V_p}^-$ is displayed as a function of $T_S$ and $h_{\rm exc}$ in the nonlinear regime. The blue tones are linked to the negative conductance (thermoelectricity), while the red area is referred to the positive one (dissipation).}
\label{Fig2}
\end{figure}

Here, we investigate thermoelectric effects in the $S-I-S_m$ junction as a function of the thermal bias and exchange field. Typical current-voltage characteristics for different values of $h_{\rm exc}$ at $T_S=0.7T_c$ and $T_{S_m}=0.01T_c$ are shown in Fig.~\ref{Fig2}(a). For $h_{\rm exc}=0$ (black solid line), the system is dissipative [$I_q(V)V>0$], and the current displays sub-gap peaks at $V=\pm V_p=\pm|\Delta(T_S)-\Delta(T_{S_m})|/|e|$. It is enough a weak exchange field ($h_{\rm exc}=0.1\Delta_0$, magenta solid line) to observe the splitting of the peaks at $V=\pm V_p^{\mp}=\pm |\Delta(T_S)-\Delta(T_{S_m})\pm h_{\rm exc}|/|e|$. For larger values of the exchange field ($h_{\rm exc}\gtrsim 0.2\Delta_0$), a thermoelectric power is generated [$I_q(V)V<0$]. Thermodynamic analysis~\cite{MarchegianiPRL} shows that in a thermally biased non-equilibrium system the generation of thermopower in the junction is possible, which corresponds to an Absolute Negative Conductance (ANC) \cite{Supp}. The exchange field generates spontaneously a thermo-spin effect (discussed in more details below). The spin-splitting in $S_m$ is fundamental to activate this effect when $\Delta_{S,0}=\Delta_{S_m,0}$. Indeed, the spin-splitting in the bottom electrode DoS reduces effectively the gap, and, at the same time, localizes purely spin-polarized states between $\Delta_{S_m} - h_{\rm exc}$ and $\Delta_{S_m} + h_{\rm exc}$. These combined mechanisms determine a spontaneous spin-Seebeck effect, leading to thermoelectricity and ANC, that is $G(V)=I_q(V)/V<0$. For $V\to 0$, the conductance can be approximated by
\begin{equation}
\frac{G_{0}}{G_T} \simeq -\Delta_{0}^2 \sum_{\sigma}\! \int_{0}^{\infty} \!\!\!\!\!\!\!dE \frac{N_{S}(E) f_{S_m}(E,T_{S_m})}{[(E+i\Gamma+\sigma h_{\rm exc})^2-\Delta_{0}^2]^{3/2}}.
\label{Eq:G0}
\end{equation}
See for instance the dashed violet line ($I_q=G_0 V$) for $h_{\rm exc}=0.2\Delta_0$ in Fig.~\ref{Fig2}(a). Eq.\ref{Eq:G0} holds only for $\Delta(T_S)>\Delta(T_{S_m})-h_{\rm exc}$, that is when thermoelectricity appears. Figure~\ref{Fig2}(b) displays $G_0$ computed through numerical differentiation of $I_q$ as a function of ~$T_{S}/T_{c}$ and $h_{\rm exc}/\Delta_0$. We can distinguish thermoelectric ($G_0<0$, blue) and dissipative ($G_0>0$, red) areas. For a fixed $h_{\rm exc}$, thermoelectricity is present only in a limited range of $T_S$. In particular, the maximum value of $T_S$ providing a thermoelectric effect is due to the condition $\Delta(T_S)>\Delta(T_{S_m})-h_{\rm exc}$ (dashed white line). This constrain corresponds to the requirement to have the hot electrode with the largest "effective" gap~\citep{MarchegianiPRL, MarchegianiPRB}.\\
The thermoelectric power is typically maximum at the internal peak voltages $V=\pm {V_p^-}=\pm |\Delta(T_S)-(\Delta(T_{S_m})- h_{\rm exc})|/|e|$. Being $\Delta(T_S)<\Delta(T_{S_m})$, we note that $|{V_p^-}|$ can be increased by raising $h_{\rm exc}$. As a consequence, the generated maximum thermopower $-I_q(V_p^-)V_p^-$ is expected to increase accordingly. Note that $h_{\rm exc}$ cannot be freely increased, due to the Chandrasekhar-Clogston limit ($h_{\rm exc} \le \Delta_0/\sqrt{2}$)~\citep{Clogston, Chandrasekhar, Maki}.\\ A typical thermoelectric figure of merit is the Seebeck voltage ($V_S$), which represents the bias that stops the thermocurrent $I_q(V_S)=0$ [see the orange circles in Fig.~\ref{Fig2}(a)]. $|V_S|$ grows monotonically with $h_{\rm exc}$, as shown in Fig.~\ref{Fig2}(c) for different values of $T_S$, and its maximum is limited again by the Chandrasekhar-Clogston limit \cite{Supp}. Moreover, $|V_S|$ increases by lowering $T_S$, which corresponds to a \emph{decrease} of the thermal gradient. This odd behavior shows that thermoelectricity is purely nonlinear ~\citep{MarchegianiPRL}, and notably different from thermoelectricity in linear regime \cite{Ozaeta}.
Another significant figure of merit is the ratio between  the maximum thermoelectric current and the corresponding voltage, i.e., $G^{\rm max} \sim I_q^{\pm}(V_p^-)/V_p^-$ [see aquamarine dotted line for $h_{\rm exc}=0.4\Delta_0$ in Fig.~\ref{Fig2}(a)].
This quantity plays a crucial role when the thermoelectric element is connected to a load. More precisely, $-G^{\rm max}$ represents the maximal conductance of the load supported by the thermoelectric junction, such as no net thermopower can be generated if the load conductance is bigger than $-G^{\rm max}$ ~\citep{MarchegianiPRB, Supp}.
Figure ~\ref{Fig2}(d) shows $G^{\rm max}$ as a function of $h_{\rm exc}$ and $T_S$. We can identify again thermoelectric ($G^{\rm max}\!<0$, blue) and dissipative ($G^{\rm max}>0$, red) regimes. We find that the temperature range where the junction is thermo-active widens by increasing $h_{\rm exc}$. Indeed, for a given $h_{\rm exc}$, the maximum value of $T_S$ is still limited by the above-mentioned relation for $G_0$ [dashed white line of Fig.\ref{Fig2}(b)-\ref{Fig2}(d)]. For low temperatures, thermoelectricity disappears arising from the nonlinear nature of the effect in temperature~\footnote{The lower limit in temperature is nonuniversal being related to the chosen $\Gamma$ parameter.}. Some other differences between Fig.~\ref{Fig2}(b) and Fig.~\ref{Fig2}(d) are discussed in more details in \cite{Supp}.\\
%%%%%%%%%%%%%%%%%%%%%%%%%%%%%%%%%%%%%%%%%%%%%%%%%%%%%%%%%%%%%%%%%%%%%%
\section{Spin thermoelectric effect}
Here, we investigate the thermo-spin current, which represents another peculiar consequence of spin symmetry breaking in the setup considered.\\
\begin{figure}[htp!]
\includegraphics[width=1\columnwidth]{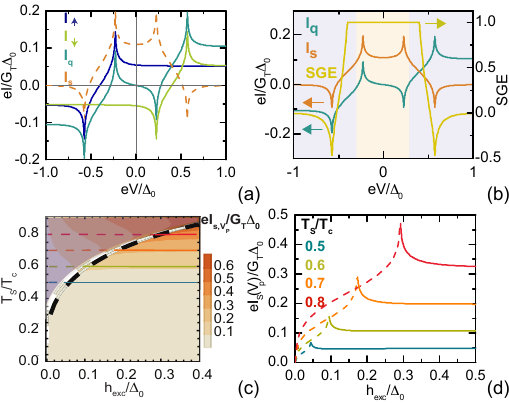}
\caption{(a) The charge current ($I_{q}$) is separated into up ($I_{\uparrow}$) and down ($I_{\downarrow}$) spin components for $h_{\rm exc}=0.4\Delta_0$. Only one spin component ($I_{\downarrow}$ for $V>0$, $I_{\uparrow}$ for $V<0$) generates a thermo-active peak. The spin current ($I_s$) is reported in the dashed orange line. (b) The IV characteristics of the charge current ($I_{q}$) and spin current ($I_{s}$) and the curve of the spin current generation efficiency ($SGE$) are displayed for $h_{\rm exc}=0.4\Delta_0$. The yellow area between the Seebeck voltages ($\pm V_s$) highlights the thermo-active spin current generation, while in the violet area the system is dissipative behaving as a spin-filter. (c) The spin current evaluated at $+{{V}_p}^-$ is reported as a function of $T_S$ and $h_{\rm exc}$. Fixing $T_S$ (colored lines), the maximal value of $I_s$ is reached for the optimal value of $h_{\rm exc}$ such as $\Delta_S=\Delta_{S_m}-h_{\rm exc}$ (dashed white line). With a black dashed line, we show the approximation of $h^t_{\rm exc}$. The violet transparent zone highlights the dissipative behavior of the system.
(d) The maximum spin current, evaluated at the matching peak voltage ($+{{V}_p}^-$), corresponding to the colored cuts of the previous panel(c) (dashed lines for the dissipative regime and solid lines for the thermoelectric regime), is shown as a function of $h_{\rm exc}$ for different values of $T_S$.}
\label{Fig3}
\end{figure}
Figure~\ref{Fig3}(a) shows typical charge current ($I_q$, aquamarine line) and spin current ($I_s$, orange dashed line) as a function of $V$ in the presence of thermoelectricity. In addition, we consider the two spin components of $I_q$, i.e.,  $I_{\uparrow}$ (spin up, blue line) and $I_{\downarrow}$ (spin down, light green line). At $V=0$, $I_q(V=0)=0$, due to PH symmetry. Still, the two spin components are finite, and exactly opposite, since $h_{\rm exc}\neq 0$. Therefore, one of the spin components is thermoactive [in Fig.~\ref{Fig3}(a) $I_{\uparrow}(V)V<0$ for $V<0$, $I_{\downarrow}(V)V<0$ for $V>0$], while the other one is dissipative. Thus, the system produces spin-polarized thermoelectricity when the thermoactive component is larger than the dissipative one. Interestingly, the Seebeck voltage represents the bias
where the thermoactive and dissipative components compensate, i.e., $I_{\uparrow}(+V_S)=-I_{\downarrow}(+V_S)$ such as $I_q(\pm V_S)=0$, thus obtaining a purely spin current. Moreover, the maximum spin current is located at the internal matching peaks ($\pm {{V_p}^-}$). \\
Hence, it is convenient to introduce a new figure of merit, which compares the
spin current generated by the system with respect to all the carriers moving across the junction. We define the Spin current Generation Efficiency as $SGE=I_S/(|I_{\uparrow}|+|I_{\downarrow}|)$. Note that when the two spin components flow in opposite directions, $|I_s|>|I_q|$ and the system is globally thermo-active [$I_q(V)V<0$], necessarily $SGE=1$. Figure~\ref{Fig3}(b) displays $SGE$ as a function of $V$, for the curves in Fig.~\ref{Fig3}(a). For $|V|>V_S$, where the junction is dissipative (violet area), we can distinguish two behaviors.
The first is characterized by $SGE=1$ with a thermoactive spin component but lower than the dissipative one,i.e., no charge thermopower is generated. In the second case, $SGE<1$. Here, both spin components are dissipative and the total charge current flows in the same direction of the bias. The current is still spin-polarized, but $|I_s|\le |I_q|$ similarly to a spin-polarizer \citep{Ozaeta, Linder, BergeretColloquium}. For large $|V|$, the total current is almost independent of the spin-splitting, and the two components give the same contribution, leading to $I_s\to 0$.\\
In Fig.~\ref{Fig3}(c), we analyse the maximal spin current evaluated at the peak $I_s(\pm V_p^-)$ as a function of $h_{\rm exc}$ and $T_S$. At fixed $h_{\rm exc}$, $I_s$ increases with the thermal bias $\delta T = T_S-T_{S_m} \sim T_S$. By contrast, for a given thermal gradient ($\delta T$, colored cuts), the spin current is non-monotonic in $h_{\rm exc}$, reaching the maximum value for $h_{\rm exc}=\Delta(T_{S_m}) -\Delta(T_S)$. This condition coincides with the threshold value of the exchange field [$h_{\rm exc}^t(T_S)$] to generate thermoelectricity, which can be estimated as $h^t_{\rm exc}=\Delta\left(1-\tanh\left[1.74\sqrt{(T_c/T_S)-1}\right]\right)$ \cite{Supp}. Its approximate value is displayed in Fig.~\ref{Fig3}(c) with a black dashed line. In particular, for a fixed $T_S$, spin-thermoelectricity occurs only for $h_{\rm exc}\geq h_{\rm exc}^t$ (solid part of the colored cuts), otherwise, it is dissipative (dashed part). 
Finally, Fig. \ref{Fig3}(d) shows $I_s(\pm V_p )$ as a function of $h_{exc}$ for selected values of $T_S$. In particular, $I_s(V_p^-)$ grows with $T_S$, reaching its maximum at $h^t_{\rm exc}$. We observe that the maximum value of the thermocurrent is of the order of $I^{max}_s \sim 0.5 G_T \Delta_0/e$. Assuming the aluminium gap of $ \Delta_0= 200~ \mu eV$ and tunneling conductance of $G_T=0.1~ \rm{mS}$, we expect spin currents of the order of 10 nA. Spin currents cannot be easily detected without further elaborating on the design. Possible methods are measuring spin accumulation phenomena or employing spin-filtering.\\
%%%%%%%%%%%%%%%%%%%%%%%%%%%%%%%%%%%%%%%
\section{Conclusions}
In summary, we discussed the nonlinear thermo-spin effect, generated in a thermally-biased $S-I-S_m$ Josephson junction in the presence of an exchange interaction. This effect is generated by the spontaneous particle-hole symmetry breaking activated by spin-splitting in $S_m$. By exploiting the two spin current components that the system naturally drives in opposite direction, we observe the coexistence of one spin thermo-active with one dissipative component. Notably, the spin thermoelectricity is relevant when the thermoactivated component dominates, thus obtaining $|I_s|>|I_q|$. Our results suggest interesting applications in thermoelectricity \citep{Benenti}, low-dissipative and thermoactive spintronics \citep{ReviewSpintronics, Linder}, and radiation detection \citep{Geng}.\\
\section{Acknowledgments}
We acknowledge the European Research Council under Grant Agreement No. 899315-TERASEC, and  the  EU’s  Horizon 2020 research and innovation program under Grant Agreement No. 800923 (SUPERTED) and No. 964398 (SUPERGATE) for partial financial support. The Royal Society through the International Exchanges between the UK and Italy (No. IEC R2 192166), and the SNS-WIS joint lab QUANTRA funded by the Italian Ministry of Foreign Affairs and International Cooperation.

%\bibliographystyle{apsrev4-1}
%\bibliography{references.bib}

%%%%%%%%%%%%%%%%%%%%%%%%%%%%%%%%%%%%%%%%%%%%%%%%%%%%%%%%%%%%%%%%%%%%%%%%%%%%%%BIBLIOGRAPHY%%%%%%%%%%%%%%%%%%%%%%%%%%%%%%%%%%%%%%%%%%%%%%%%%%%%%%%%%%%%%%%%%%%%%%%%%%%%%%%%%%%%%%%%%%%%%%%%

%%%%%%%%%
%%%%%%%%%%%%% SUPPLEMENTARY INFORMATION %%%%%%%%%%%%%%
%%%%%%%%%%%%%%%%%%%%%%%%%%%%%%%%%%%%%%%%%%%%%%%%%%%%%%

\clearpage
\onecolumngrid
\setcounter{figure}{0}
\setcounter{equation}{0}
\setcounter{section}{0}
\renewcommand\thefigure{S\arabic{figure}}
\renewcommand\theequation{S\arabic{equation}}
\section*{Supplementary Information}

\section{SYMMETRIES OF THE CHARGE CURRENT AND THE SPIN CURRENT}
We discuss the symmetries of the charge current ($I_q$) and the spin current ($I_s$), which directly derive from the particle-hole (PH) symmetry of the superconducting density of states (DoS). As discussed in the main text, the spin-$\sigma$ (with $\sigma = \pm$) DoS of the $\alpha$ superconductor (with $\alpha = S, S_m$) in the junction is \cite{Tinkham2004}
\begin{equation}
N_{\alpha \sigma}(E_\alpha)= \left |\Re \!\! \left[\frac{E_{\alpha}+ \sigma h_{\alpha} + i\Gamma}{\sqrt{(E_{\alpha}+\sigma h_{\alpha}+ i\Gamma)^2-{\Delta_{\alpha}}^2}}\right]\!\!\right |,
\label{Eq:1DOS}
\end{equation}
where $h_\alpha$ is the exchange field, $E_{\alpha}=E-\mu_{\alpha}$ is the quasiparticle energy measured with respect to the chemical potential $\mu_\alpha$, and $\Delta_\alpha$ is the self-consistent gap for the $\alpha$-lead, $\Gamma$ is the Dynes parameter. The full DoS is given by the sum $N_{\alpha}(E)=\sum_{\sigma=\pm}N_{\alpha\sigma}(E)/2$. In the presence of an exchange (or Zeeman) field  $h_{\rm exc}$, even if the spin degeneracy of $S_m$ is broken, the superconducting state of $S_m$ still satisfies the PH symmetry, which implies the symmetry $N_{\alpha\sigma}(E_\alpha)=N_{\alpha\bar{\sigma}}(-E_\alpha)$, where $\bar{\sigma}=-\sigma$. Anyway, the PH symmetry of the BCS superconductors is reflected in the BCS DoS by its even symmetry with respect to the energy, i.e. $N_\alpha(E_\alpha)=N_\alpha(-E_\alpha)$. \\
In the following, we assume to have the exchange field \emph{only} in the bottom lead, and consequently $h_S=0$, as discussed in the setup $S-I-S_m$, shown in Fig.1 of the main text. Correspondingly, the current for the $\sigma$-spin component becomes
\begin{equation}
\label{eq:Isigma}
I_{\sigma}(V)= \frac{G_T}{2e}\int_{-\infty}^{\infty}\!dE~N_{S\sigma}(E-eV) N_{S_m\sigma}(E,h_{exc}){F}_{SS_m}(E,V),
\end{equation}
as reported in the main text. In the above expression, ${F}_{SS_m}(E,V)=f_S(E-eV,T_S)-f_{S_m}(E,T_{S_m})$ is the difference between the Fermi-Dirac quasiparticle distributions of two electrodes $f_\alpha(E,T_\alpha)=[1+\exp(E/k_B T_\alpha)]^{-1}$. Note that the PH symmetry of the Fermi function with respect to the energy gives $f_\alpha(E)=1-f_\alpha(-E)$ and determines also the symmetry ${F}_{SS_m}(-E,V)=-{F}_{SS_m}(E,-V)$. By changing the integration variable $E\to-E'$ in Eq.~\eqref{eq:Isigma} and using the symmetries of $N_{\alpha\sigma}(E)$ and ${F}_{SS_m}(-E,V)$, we obtain
\begin{equation}
I_{\sigma}(V)= -\frac{G_T}{2e}\int_{-\infty}^{\infty}\!dE'~N_{S\sigma}(E'+eV) N_{S_m\bar{\sigma}}(E',h_{exc}){F}_{SS_m}(E',-V)=-I_{\bar{\sigma}}(-V),
\end{equation}
where the second identity is obtained by direct comparison with  Eq.~\eqref{eq:Isigma}. 
By exploiting the symmetry of $I_\sigma$, it is easy to show that charge ($I_q=\sum_{\sigma =\pm}I_\sigma$) and spin ($I_s=\sum_{\sigma =\pm}\sigma I_\sigma$) currents are odd and even functions in $V$, respectively. For instance, in the case of the symmetry with respect to the voltage bias $V$, one gets
\begin{equation}
\begin{split}
I_q(-V)&=-\sum_{\sigma=\pm}I_{\bar{\sigma}}(V)=-I_q(V)\\
I_s(-V)&=-\sum_{\sigma=\pm}\sigma I_{\bar{\sigma}}(V)=\sum_{\sigma=\pm}\bar{\sigma} I_{\bar{\sigma}}(V)=I_s(V)
\end{split}
\end{equation}
Similarly, one can easily show that charge (spin) current is even (odd) with respect to the exchange field $I_q(V,h_{exc})=I_q(V,-h_{exc})$ [$I_s(V,h_{exc})=-I_s(V,-h_{exc})$].\\
%%%%%%%%%%%%%%%%%%%%%%%%%%%%%%%%%%%%%%%
\section{Evaluation of $V_p^-$ and the $h_{exc}$ thermoelectrical threshold}
In the presence of thermoelectricity, and assuming that the critical temperature $T_c$ of the superconductors is $T_c \gtrsim T_S\gg T_{S_m} $, the value of the voltage bias where the thermoelectric current is maximum reads
\begin{equation}
   eV_p^-=\Delta(T_S)-\left[\Delta(T_{S_m}, h_{exc})-h_{exc}\right].
   \label{Eq:Eq6}
\end{equation}
For low temperatures ($T_{S_m} \ll T_c$) and low exchange fields, we can approximate $\Delta(T_{S_m},h_{exc})\approx \Delta_0$. We can estimate the maximum value of the peak voltage ($V_p^{-,max}$) by considering the maximum exchange field allowed, i.e., $h_{exc}\leq h_{exc}^{max}=\Delta_0/\sqrt{2}$, given by the Chandrasekhar-Clogston limit \citep{Clogston, Chandrasekhar}. We obtain
\begin{equation}
    eV_p^{-,max}\approx \Delta_0(\tanh[1.74\sqrt{T_c/T_S-1}]-0.29),
\end{equation}
where we approximate $\Delta(T_S)=\Delta_0 \tanh[1.74\sqrt{T_c/T_S-1}]$ \cite{Tinkham2004}, which overall differs only up to a few percent from the results obtained through self-consistent calculation of $\Delta_T$. This limit provides also a rough estimation of the Seebeck voltage values, being $V_S\gtrsim V_p^{-,max}$.
By using a similar approximation, we can compute the minimum threshold value for $h_{\rm exc}$ necessary to generate thermoelectricity in the system. We observe that the matching peak is located at zero bias, i.e., $V_p^{-}=0$, when the system switches from dissipative to thermoactive behavior. Thus, using again the Eq.~\eqref{Eq:Eq6} we find
\begin{equation}
h_{\rm exc}^t=\Delta_B(T_{S_m},h_{\rm exc}^t)-\Delta_T(T_S,h_{\rm exc}=0).
\end{equation}
Using previous approximations, we find
\begin{equation}
\Delta_0(\tanh[1.74\sqrt{T_c/T_S-1}]-[1-h_{\rm exc}^t/\Delta_0])=0
\end{equation}
and the threshold $h_{exc}^t$ becomes
\begin{equation}
h_{\rm exc}^t=\Delta_0\left[1-\tanh\left(1.74\sqrt{\frac{T_C}{T_S}-1}\right)\right].  
\end{equation}
The relation implies that by increasing the temperature ($T_S$), $h_{\rm exc}^t$ rises as shown by the Seebeck voltage $V_S$ of Fig. 2(c) of the main text.
From the previous results, we expect that the Seebeck voltage $V_S$ depends linearly on $h_{\rm exc}$ and $h_{\rm exc}^t(T_S)$. We can provide a very rough estimation of the Seebeck voltage with  
\begin{equation}
e V_S^-\approx  h_{\rm exc}- h_{\rm exc}^t(T_S)=h_{\rm exc}-\Delta\left[1-\tanh\left(1.74\sqrt{\frac{T_C}{T_S}-1}\right)\right].  
\end{equation}

\section{The negative conductance in linear and  non linear regime}

\begin{figure}[H]
\includegraphics[width=1\columnwidth]{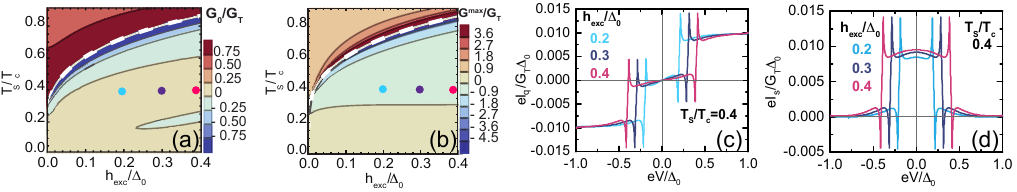}
\caption{(a) The negative conductance $G_0$ is reported as a function of $T_S$ and $h_{\rm exc}$. Blue tones (red tones) show a thermoelectric (dissipative) behavior. The colored points are referred to the IV characteristics of Fig.~\ref{Fig1Suppl}(c) at $T_S/T_c =0.4$ and $h_{\rm exc}$ fixed. (b) The maximal conductance $G^{\rm max}$ evaluated at $\pm V_p^-$ is reported as a function of $T_S$ and $h_{\rm exc}$. In the same way, blue tones (red tones) show thermoelectricity (dissipative regime). The colored points are referred to the IV characteristics of Fig. \ref{Fig1Suppl}(c) at $T_S$ and $h_{\rm exc}$ fixed. (c) The $I_q(V)$ characteristics of the charge current are shown at $T_S/T_C = 0.4$ for different values of $h_{\rm exc}$. (d) The $I_s(V)$ characteristics of the spin current are displayed at $T_S/T_C = 0.4$ for different values of $h_{\rm exc}$.}
\label{Fig1Suppl}
\end{figure}

Before discussing in details the absolute negative conductance in linear and nonlinear regime, we wish to recollect here some general consequence of the physical significance of a negative conductance. Firstly we observe that, as discussed in Ref.\cite{MarchegianiPRL}, the Absolute Negative Conductance (ANC) $G=I(V)/V<0$ is necessarily present when the power is produced in the system (such as in the thermopower generation), i.e. the current flows against the bias $I(V)V<0$. This is thermodinamically admissible, if there is a gradient of temperature in the junction, similar to what happen in thermoelectrical systems. Notably the system present also a Negative Differential Conductance (NDC), which can be present both in the dissipative $G>0$ and, also, in the thermoactive regime with ANC, $G<0$. 
This fact has important consequences on the electrical stability. Indeed, for a certain operating points, when ANC and NDC are present contemporaneously, an electrical instability occurs and it can be kept stable only by applying an external generator (power source) to the system. In other words, in such cases, if the junction is electrically connected to a load (dissipative element), the junction progressively increases the bias until the differential conductance becomes positive. In such condition, with positive differential conductance, the system may be again electrically stable. Thus, in the absence of an external sources, assuming that the junction is connected only to a passive load, the circuit naturally flows outside from the unstable electrical configuration toward a stable one, where the differential conductance is positive. This peculiar behaviour reflects the spontaneous particle-hole breaking property of the junction, which is a purely intrinsic mechanism of the junction. Note that when the load conductance $G_L$ satisfies the condition $G_L<|G^{\rm max}|$, this stable point can be found also in the presence of ANC. The junction, in that case, operates as a thermoelectric generator and the thermopower is dissipated in the load, as a thermoelectric engine\cite{MarchegianiPRB}. Thus, we note the rich character of the electrical behaviour of the junction. It presents thermoactive or dissipative behaviour, which can be electrically stable or unstable, depending on the internal conditions such as lead temperatures or external one due to the circuit connected to the junction.\\
In the main text, we discussed the negative conductance ($G_0$) in the linear regime and the maximal conductance ($G^{\rm max}$) in the nonlinear regime as a function of $h_{\rm exc}$ and $T_S$ [Fig.~\ref{Fig1Suppl}(a-b)]. We observed the condition for thermoelectricity (blue tones) and its limits in temperature at fixed $h_{\rm exc}$.\\
Here, we focus on the differences that appear between the linear and nonlinear regimes. In particular, observing the central area of both figures ($T_S/T_c \sim 0.4$), the system is dissipative $G_0>0$ around $V\sim 0$ [Fig.~\ref{Fig1Suppl}(a)], while it exhibits thermoelectric effects $G^{\rm max}<0$ at the matching peak $V=\pm V_p^-$ [Fig.~\ref{Fig1Suppl}(b)].\\
For clarity, we report IV characteristics of the charge current at fixed temperature ($T_S/T_C=0.4$) for different values of $h_{\rm exc}$ [Fig.~\ref{Fig1Suppl}(c)], as indicated by the colored dots in the Fig.~\ref{Fig1Suppl}(a). We can note the dissipative behavior [$I_q(V)/V>0$] around $V \sim 0$ and thermoelectricity [$I_q(V)/V <0$] around $\pm V_p^-$. This difference between linear and nonlinear in bias behavior has been reported also in absence of $h_{\rm exc}$ for nonlinear thermoelectric systems~\cite{MarchegianiPRB}.
For a more complete discussion, we show also the spin current for the same values. Note that the curves display a different functional shape with respect to the case analyzed in the main text at $T_S/T_c=0.7$ and for $h_{\rm exc}=0.4 \Delta_0$, which represents the typical behavior [Inset of Fig. 1(b)].\\
We observe that the reverse behavior is also observed. Indeed, for high values of $h_{\rm exc}$ at $T_S/T_c \sim 0.2$ , the thermoelectricity is present only in the linear regime ($V \sim 0$) [Fig. \ref{Fig1Suppl}(a)]. In this case, the thermoelectric currents are very small and strongly dependent on the Dynes parameter representing a non typical feature.

\section{Dependence of the IV curves on the values of the Dynes parameters $\Gamma$}
\begin{figure}[H]
\includegraphics[width=1\columnwidth]{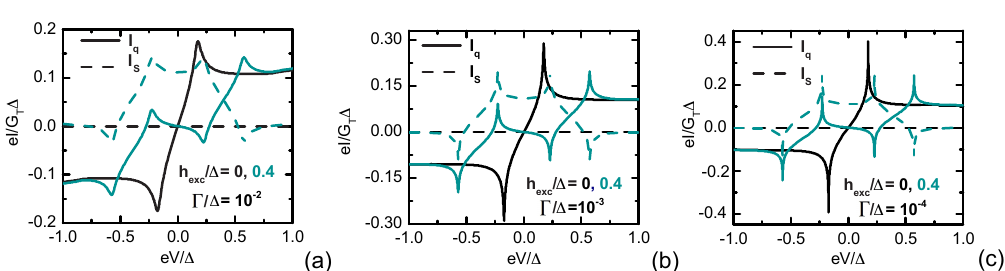}
\caption{Dependence of the charge $I_q(V)$ (solid lines) and spin current $I_s(V)$ 
(dashed lines) in the thermoelectrical regime from the variation of the Dynes parameters. The figure represents the inset of Fig.1 where the currents are taken for $h_{exc}=0$ (blue lines) or $h_{exc}/\Delta=0.4$ for different values of $\Gamma/\Delta=10^{-2}$(a), $10^{-3}$(b) and $10^{-4}$(c). Panel (b) coincides with the inset of Fig. 1.}
\label{Fig2Suppl}
\end{figure}

The phenomenological Dynes parameter introduced in the main text is usually exploited to include non-universal effects due to the junction quality \cite{Dynes} and/or interactions with the electromagnetic environments \cite{Pekola10}, which affect the quasi-particle lifetime introducing subgap states. These effects are typically observed by the presence of a subgap resistance both in normal-insulator-superconductors (NIS) and superconductor-insulator-superconductor (SIS). 
%which is also typically adopted to evaluate the parameter. 
In the paper we adopted a typical value $\Gamma/\Delta=10^{-3}$, in agreement with experimental analysis \cite{Strambini}, which have reported even lower values up to $10^{-4}$ in such hybrid ferromagnetic-superconductor junctions. \\
In Fig. \ref{Fig2Suppl}, we investigate the evolution of the charge $I_q(V)$ (solid lines) and spin $I_s(V)$ (dashed lines) current with ($h_{exc}/\Delta=0.4$, red lines) and without ($h_{exc}/\Delta=0$, blue lines) spin-splitting for three different values of the Dynes parameter $\Gamma/\Delta=10^{-2}$(panel a), $10^{-3}$(panel b) and $10^{-4}$(panel c). For comparison, we report in the panel (b) the inset of Fig. 1b of the main text.
We note immediately that, even if this parameter changes of two orders of magnitude, the general behavior of the IV characteristics is not crucially affected. The quantity mostly affected by the change of the $\Gamma$ parameter is the current value at the matching peak $V=\pm V_p$, which changes roughly of a factor two from panel (a) to panel (c). This is quite expected, since one of the most important role of $\Gamma$ parameter is to re-normalize phenomenologically the singularity at the matching peaks. Outside of that bias value, the characteristics are not crucially modified, neither the values of the NDC in the linear regime $G_0$ neither the Seebeck voltage. Thus, the spin-thermoelectric effect we reported is not crucially affected by the non-universal effects described by Dynes parameter.
However, if the quality of the junction is too low and the Dynes parameter becomes too high, it introduces too many dissipative subgap states, which can even suppress the thermoelectric effect (not shown).

\begin{comment}
%\bibliographystyle{apsrev4-1}
%\bibliography{Ref.bib}

\end{comment}

\end{document}